# Insurance Business and Sustainable Development

Dietmar Pfeifer[1] and Vivien Langen[2]

January 25, 2021

**Abstract**

In this study, we will discuss recent developments in risk management of the global financial and insurance business with respect to sustainable development. So far climate change aspects have been the dominant aspect in managing sustainability risks and opportunities, accompanied by the development of several legislative initiatives triggered by supervisory authorities. However, a sole concentration on these aspects misses out other important economic and social facets of sustainable development goals formulated by the UN. Such aspects have very recently come into the focus of the European Committee concerning the Solvency II project for the European insurance industry. Clearly the new legislative expectations can be better handled by larger insurance companies and holdings than by small- and medium-sized mutual insurance companies which are numerous in central Europe, due to their historic development starting in the late medieval ages and early modern times. We therefore also concentrate on strategies within the risk management of such small- and medium-sized enterprises that can be achieved without much effort, in particular those that are not directly related to climate change.

We start this study with a general overview of the UN sustainable development goals and their implementation in the financial sector world-wide, with a major focus on climate change aspects of investments in a lower carbon economy and economic support of underdeveloped countries that were prevailing until very recently. Although the insurance sector can be considered as a particular branch of the finance industry there are several particularities which need a separate consideration. In the first place, insurance provides a protection of individuals and companies against severe material and non-material losses. Therefore the insurance premiums must be invested safely, in particular under actual insurance regulations like Solvency II. But the insurance industry is also faced with new emerging risks due to climate change, in both the life and non-life sector. Moreover, the European development of insurance regulation has very recently focused also on other sustainability aspects than those related to climate change. We discuss this aspect of risk management in a separate section of this study. Finally, we discuss in detail appropriate strategies how small- and medium sized insurance companies in Europe can handle the new challenges of insurance supervision without too much effort. Our suggestions are mainly driven by own experiences from practice.

## 1. Introduction

In 2015, the United Nations (UN) member states adopted a far-reaching resolution with the intention to transform the world [1, p. 7]. Under the impression that the targets that were originally formulated in the Millennium Development Goals (MDGs), which were the dominating political framework from 2000 to 2015, were seemingly not completely reached, the MDGs were replaced by the 2030 Agenda for Sustainable Development [2]. The eight MDGs were replaced by seventeen Sustainable Development Goals (SDGs) that should be achieved world-wide by 2030:

1  No Poverty
2  Zero Hunger
3  Good Health and Well-Being
4  Quality Education
5  Gender Equality
6  Clean Water and Sanitation
7  Affordable and Clean Energy
8  Decent Work and economic growth
9  Industry, Innovation and Infrastructure
10  Reduced Inequalities
11  Sustainable Cities and Communities

[1] University of Oldenburg, Institute of Mathematics, Germany, and
  Chairman of the Supervisory Board of the GVO Mutual Insurance Company Oldenburg, Germany
[2] GVO Mutual Insurance Company Oldenburg, Germany, Sustainability Officer

12  Responsible Consumption and Production
13  Climate Action
14  Life below Water
15  Life on Land
16  Peace, Justice and Strong Institutions
17  Partnership

A parallel initiative to these goals for long-range sustainable economic and business activities was the development of the concept of Corporate Social Responsibilities (CSR) [3]. "More specifically, CSR for example involves fair business practices, staff-oriented human resource management, economical use of natural resources, protection of the climate and environment, sincere commitment to the local community, and also responsibility along the global supply chain" [4, p. 3]. PRI Association, an investor initiative in partnership with UNEP Finance Initiative and UN Global Compact, has introduced Environmental, Social, and Governance (ESG) issues in their principles for sustainable investments to varying degrees across companies, sectors, regions, asset classes and through time [5, p. 2]. Some examples of ESG issues are [6, p. 3]

| | |
|---|---|
| Environmental (E) | Biodiversity loss, greenhouse gas emissions, climate change, renewable energy, energy efficiency, environmental pollution, waste management, ozone depletion, changes in land use, ocean acidification, changes to the nitrogen and phosphorus cycles |
| Social (S) | Human rights, labour standards in the supply chain, child labour, workplace health and safety, freedom of association and freedom of expression, human capital management and employee relations, diversity, activities in conflict zones, controversial weapons |
| Governance (G) | Company board structure, size, diversity, skills and independence, executive pay, shareholder rights, stakeholder interaction, disclosure of information, business ethics, bribery and corruption, internal controls and risk management |

In the course of time, CSR has moved from a type of international private business self-regulation, along the lines of the UN SDGs expressed through the ESGs, to a more generally accepted source of principles and mandatory schemes at regional, national and international levels, including bilateral investment treaties and free trade agreements [7].

## 2. The role of the financial sector in sustainable development

Climate change and ESG issues have strongly influenced the finance sector – both banking and insurance – world-wide in the last decades [8]. The manner in which institutional investors approach ESG issues is gaining increased attention in particular across OECD countries. Pension funds, insurers and asset managers have to understand and respond to potential risks and opportunities arising from ESG-related factors in order to safeguard the assets that they invest on behalf of their beneficiaries and clients. At the same time, regulators must be confident that institutional investors meet the required standards of prudence and care when they include ESG considerations in their portfolio decisions [9]. This is also stressed in the reports of the Global Sustainable Investment Alliance who define sustainable investing as an investment approach that considers ESG factors in portfolio selection and management [10].

Concerning climate change aspects, the Financial Stability Board Bank for International Settlements has established a Task Force on Climate-related Financial Disclosures (TCFD), which establishes recommendations for disclosing clear, comparable and consistent information about the risks and opportunities presented by climate change. Their widespread adoption will guarantee that possible effects of climate change become regularly considered in business and investment decisions. A routine use of these recommendations will also support companies in a better demonstration of their responsibility and foresight in climate issues [11]. Nearly 200 countries agreed in December 2015 to reduce greenhouse gas emissions and accelerate the transition to a lower-carbon economy. A particular issue here are investments in alternative clean, energy efficient energy sources, for example, windmills, solar energy or water power. The expected transition to a lower-carbon economy is estimated to require around $1 trillion of investments a year for the proximate future, thereby also generating new investment opportunities. The United Nations Environment Programme (UNEP) has been supporting the idea of creating a sustainable financial system since 2014 with a purpose of mobilizing capital for sustainable development and achieving a green and inclusive economy.

A particular strategy that has been singled here out is green finance. Globally speaking, developing economies face serious challenges concerning mobilizing capital related to green investments. For these countries, one source for external capital flow is represented by foreign direct investments (FDI), which generally target projects related to energy, waste, water, or agricultural development. In addition to FDI, other sources of external capital flow are concessional loans from international financial institutions, long-term commercial debts, aid and remittances. The major aim of issuing green bonds is raising financial resources for climate change initiatives. These fixed-income instruments are generally oriented to climate-friendly activities. The performance of green bonds issued in US dollars and Euros has been superior as compared to non-green bonds [12]. A recent study by the National Bank of Belgium has investigated this topic in more detail [13]. It is interesting to notice that the authors have applied sophisticated statistical methods to obtain their conclusions. In particular, their findings were:

- No statistically significant difference was seen between the overall distribution, the mean or median of bond's asset swap spread changes on individual bond pairs.
- There are indications that the volatility of some green-bonds is lower than their non-green counterparts.
- There are indications that sustainable investments like green bonds are potentially more immune to systemic crises.

Sophisticated econometric statistical methods were also the basis of the recent paper [12]. For their analysis, the authors used the variables domestic credit from banks and domestic credit from the financial sector in USA, Canada and Brazil. Gross domestic product (GDP) was used as a proxy for sustainability of economic growth, along with $CO_2$ and $N_2O$ emissions, which are caused by manufacturing, agriculture, the use of forests and fisheries. According to their findings, bank credit is insufficient to achieve green financing. For the purpose of increasing economic growth and reducing global warming, the financial sector should assume a bigger role in increasing green investments. Their results show that the level of domestic credit within financial sectors contributes to green financing, while $CO_2$ emissions remain a challenge for reaching the 1.5° C target.

The use of science-based methods in the judgement of climate related risks is also stressed in the Technical Supplement by the TCFD [14]. In general, the most significant effects of climate change are probably emerging over a medium- to long-term time horizon, but their precise timing and magnitude are uncertain. This uncertainty induces challenges in understanding the potential effects of climate change on business, strategies, and financial performance. It is, therefore, important to investigate how climate-related risks and opportunities may potentially evolve and how they affect business under different conditions. One way to assess such implications is through the use of scenario analysis.

Scenario analysis is a well-established method for developing input to strategic plans in order to enhance plan flexibility or resiliency to a range of future states. The use of scenario analysis for assessing climate-related risks and opportunities and their potential business implications, however, is relatively recent. Given the importance of forward-looking assessments of climate-related risk, the Task Force believes that scenario analysis is an important and useful tool for an organization to use, both for understanding strategic implications of climate-related risks and opportunities and for informing stakeholders about how the organization is positioning itself in light of these risks and opportunities. It also can provide useful forward-looking information to investors, lenders, and insurance underwriters.

This topic has also been addressed in a Working Group of 16 banks piloting the TCFD Recommendations under a UNEP Finance Initiative, with a special emphasis on credit risk [15]. The physical aspects from a changing climate were accompanied by a follow-up study [16].

In the insurance and reinsurance industry, these aspects have already been the basis of judgements on the effects of natural catastrophes like windstorm, hailstorm or flooding on insured portfolios for a long time [17].

Another central aspect in this discussion is the public disclosure of sustainable business activities. In order to implement and internalize the sustainability by businesses, it should, first of all, be traceable and measurable. This is possible through sustainability reporting. Sustainability reports show that there are differences among countries and even among sectors. In developed countries such as the USA, the UK, and Australia, such reports contain – besides climatic and environmental aspects – qualitative information in social and governance areas such as number of employees, salary and bonuses, and employee training [18].

## 3. The role of the insurance sector in sustainable development

Insurance companies - life insurers as well as providers of property and casualty, health, and financial coverage - perform important economic functions and are big players in financial markets. They enable economic agents to diversify idiosyncratic risk, thereby supplying the necessary preconditions for certain business activities. They are a major source of long-term risk capital to the real economy, and are among the largest institutional investors [19]. The overall assets invested by the insurance sector in 2018 were more than 32 trillion US $ [20].

Although the insurance sector is generally seen as a part of the financial sector there are, however, some peculiarities. Essentially, insurance is the process of an exchange of unpredictable financial risks (whether for individuals or for institutions) against a fixed monetary premium. The statistical basis for insurance to work economically sufficient is the famous law of large numbers discovered by Jakob I Bernoulli in the late 17$^{th}$ century [21]. Therefore sophisticated actuarial risk models and elaborate statistical calculations are a fundamental basis of insurance [22]. While the actuarial processes for insurance have been in continuous development since early on, it really took until the second half of the twentieth century for a modern theory of insurance economics to emerge [23]. The central idea here is the concept of risk diversification, which also plays an essential role in insurance regulation. Since in particular life insurance requires an utmost degree of safety in financial asset investments, governmental regulation is of great importance here; and has been set to work in almost all developed countries over the world. In Europe, this was accomplished by the Solvency II project finalized in 2016 [24].

The investment strategy (asset management) of insurance companies is limited by regulations and driven by a number of internal and external factors [25].

Insurers must invest conservatively. They must ensure that they remain solvent throughout and are able to make their payouts to the policyholders with the highest probability at any time. Insurers have a fiduciary obligation to keep or augment the value of their 'policyholder' assets. This poses constraints on the industry's investment strategies.

Furthermore, insurance regulators impose risk-based capital charges on investments to ensure adequate capital levels to cover insurers' liabilities; the riskier the investment, the higher the capital charge. These vary by country and region. It is important to note that different lines of business are exposed to different risks. That is why financial risks associated with assets and liabilities are managed differently by life and non-life insurers. Specifically,

- Life insurers are typically 'buy and hold' investors. They aim to generate predictable and stable income to match cash flows of long-dated and generally predictable liabilities. Life insurance contract duration can range from ten years to several decades, involving payout patterns of 20 to 30 years [25]. Life insurers are deeply concerned about the asset-liability mismatch, with interest rate risk being a key issue.

- Non-life insurers are geared towards more liquid investments with shorter time horizons, typically one to three year in duration [26]. However, in some instances (e.g. asbestos-related), claims are paid out many years later, exposing them to interest rate risk.

The discussion of sustainable developments in the finance sector as outlined in the preceding section has, of course, also reached the insurance sector. Firstly, one can distinguish between sustainability risks and opportunities on the asset side and on the insurer's liability side [3].

Major issues that can potentially arise from sustainability risks on the asset side include credit risk, market risk, liquidity risk, insurance risk, strategic risk and reputational risk. The German supervisory authority BaFin [26, p.18] provides the following ostensive examples:

i) Credit risk/counterparty default risk: A credit institution providing a loan to an entity with a business model that is significantly damaged by political decisions on ESG issues (such as a $CO_2$ charge).

ii) Market risk: A pension fund or investment fund could be invested in companies which do not demonstrate sustainable management or use the invested monies for transition towards sustainability. An abrupt change in market sentiment (e.g. to reflect the cost of regulatory measures) might lead to declines in value.

iii) Liquidity risk: After a catastrophic flood, tens of thousands of clients withdraw money from their accounts at a regional credit institution in order to finance damage repairs. The credit institution has to sell a high level of assets to cover these outflows.

iv) Insurance risk: Homeowners' insurance claims rise as a result of storms, floods or hail. Business interruption insurance claims may also rise. The increasing intensity and/or frequency of such events should be appropriately reflected in the assessment of technical provisions or premium risk. In this context it is also worth considering that insurance undertakings may be affected by the same sustainability risk on both the asset and the liability side.

v) Strategic risk: A credit institution specialized in financing coal mining loses the basis of its business.

vi) Reputational risk: An investment fund is invested in a clothing factory owned by a well-known brand in East Asia. The building burns down as a result of inadequate national safety standards, hundreds of workers die, reports circulating in the media name the investor. The sale of allegedly sustainable financial products (known as greenwashing) to those seeking ESG-compliant investments may also represent a reputational risk.

Important issues that can potentially arise from sustainability risks on the liability side include natural catastrophes due to windstorm, hailstorm and flooding. Beyond insured losses from physical climate damages, climate trends and shocks can cause far-reaching economic disruptions. The insurance "protection gap" for weather related losses remains significant, with roughly 70% of losses uninsured. This leaves significant burden on households, businesses, and governments. Uninsured losses arising from physical risks may have cascading impacts across the financial system, including impacts on investment companies and banks. Likewise, the availability of insurance – or risk of uninsurability due to high physical risk profiles – can have significant impacts on the performance of credit and investment across the economy (including, for instance, mortgage lending) [27]. Historically, insured risks from natural disasters were to a great extend covered by world-wide operating reinsurers with a high grade of global diversification. In the recent years, new financial products were created shifting insurance risks to the financial market, e.g. cat bonds or other climate related derivatives [28]. However, as the severity and frequency of significant natural disasters increases, the availability and cost of reinsurance cover for weather-related risks may become prohibitive for smaller insurers in certain markets – potentially leading to a reinsurance gap [27].

Another possible threat is a rise in mortality due to climate change. Extreme high air temperatures contribute directly to deaths from cardiovascular and respiratory disease, particularly among elderly people. In the heat wave of summer 2003 in Europe, for example, more than 70 000 excess deaths were recorded. High temperatures also raise the levels of ozone and other pollutants in the air that exacerbate cardiovascular and respiratory disease. [29]. Life and health insurers are in many cases just beginning to explore the impacts of climate factors on their underwriting portfolios. The potential impacts of climate change on mortality – in particular due to extremes in weather events like excessive heat – are coming into the focus of actuarial associations, who are exploring the matter in relationship to insurance, annuity and pension programmes [27].

Besides the pure monetary aspects of climate change risks and their management, also other ESG criteria have recently come into the focus of the insurance industry and their supervisors. An important lesson learned is the need for financial supervisory authorities, as well as the supervised companies, to be deeply engaged in efforts that incorporate ESG risks into their business. This requires a profound change of mind-set within institutions. In order to attain this engagement, it is very important for supervisors to raise awareness of ESG issues through provision of information, guidance, and capacity building [27].

## 4. The European way for the insurance sector in sustainable development

In 2018, the European Insurance and Occupational Pensions Authority (EIOPA), received a request from the European Commission for an opinion on sustainability within Solvency II, with a particular focus on aspects relating to climate change mitigation [30]. According to EIOPA's understanding, the term "climate risks" aims to include all risks stemming from trends or events caused by climate change, i.e., *climate change-related risks*. This encompasses extreme weather events, including natural catastrophes, but also more general climate trends such as a general rise in temperature, sea level rise, or climate-related forced migration that could affect (re)insurance activity. Concerning the impact of climate change-related risks on non-life, health and life insurance, EIOPA tried to collect information from non-life (re)insurance business. This initial step was motivated by the consideration that non-life lines of business may be affected by climate change effects over a shorter time period than the life and health business. In addition, EIOPA started to collect additional evidence on the impact of climate change related risks on the morbidity and mortality risks through a public consultation. An integration of sustainability risks in Pillar 1 of Solvency II has to take account of capital requirements within the overall Solvency II framework which aims to ensure that undertakings can survive severe unexpected shocks (losses) and still meet their obligations to policyholders over a one-year period (Article 101(3) of the Solvency II Directive [32]). The Solvency II Directive expresses this as the ability to withstand shocks with a 1 in 200 probability within this one-year time horizon.

Capital requirements in Solvency II are calibrated based on a one-year time horizon, while sustainability risks are generally considered to be long-term risks. In particular, climate change-related risks are expected to emerge over a longer time horizon which presents practical challenges for integrating them in the current Pillar 1 capital requirements.

Further, specifically for traditional non-life business, the insurance cover period (during which undertakings are liable for claims that occur) just spans the next 12 months, at the end of which, undertakings can theoretically adjust the pricing for the future, based on claims experience. This repricing is, in particular, enabled by the fact that the uncertainty on the final amount of natural catastrophe claims is limited, as they are usually settled within one year after their occurrence.

Unfortunately, market participants tend to believe that they have time to adapt their investment strategy within the next 10 to 20 years, and thus firms have limited incentives to consider climate change risks, in particular, transitions risks, in their asset portfolio today. This behaviour refers to the so-called "tragedy of the horizon" coined by Mark Carney [8].

Accompanying the aforementioned aspects, the European Commission has initiated a Taxonomy Regulation (TR), agreed at the political level in December 2019, which was intended to create a legal basis for the EU Taxonomy, published as a directive in 2020 [33]. As explained in the final Report of the Technical Expert Group on Sustainable Finance (TEG) [34] the TR sets out the framework and environmental objectives for the Taxonomy, as well as new legal obligations for financial market participants, large companies, the EU and Member States. The EU Taxonomy is a tool to help investors, companies, issuers and project promoters navigate the transition to a low-carbon, resilient and resource-efficient economy. The TR will be supplemented by delegated acts which contain detailed technical screening criteria for determining when an economic activity can be considered sustainable, and hence can be considered Taxonomy-aligned.

Consistent with the EU Action Plan on Financing Sustainable Growth, finance is a critical enabler of transformative improvements in existing industries in Europe and globally. The OECD estimates that, globally, EUR 6.35 trillion a year will be required to meet Paris Agreement goals by 2030. Public sector resources will not be adequate to meet this challenge, and mobilization of institutional and private capital will be necessary. [34].

A part of these reflections has, meanwhile, also found entry in the forementioned Taxonomy Directive, in particular, in Articles 9 and 10 [33]. As environmental objectives, the following topics are considered:

(a) climate change mitigation;
(b) climate change adaptation;
(c) the sustainable use and protection of water and marine resources;
(d) the transition to a circular economy;
(e) pollution prevention and control;
(f) the protection and restoration of biodiversity and ecosystems.

European insurance companies will be strongly affected by these political measures in the future, especially concerning their asset management. It will, however, be difficult to judge which investments are truly Taxonomy-aligned. E.g., in Solvency II, government and related bonds are considered to be the safest investment in Pillar I, but governments typically also engage in the armaments industry or fossil energy, like brown and stone coal mining in Germany, contradicting, in part, the above topics.

Recently, further ESG aspects other than mere climate change risks, have come into the focus of European insurance supervisors. For instance, the German supervisory authority BaFin compiles the following ESG topics as specific examples to be considered in the future by European insurance companies [26, p. 13]:

| | |
|---|---|
| Environmental (E) | Climate mitigation; adjustment to climate change; protection of biodiversity; the sustainable use and protection of water and maritime resources; the transition to a circular economy, the avoidance of waste, and recycling; the avoidance and reduction of environmental pollution, the protection of healthy ecosystems; sustainable land use |
| Social (S) | Compliance with recognised labour standards (no child labour, forced labour or discrimination); compliance with employment safety and health protection; appropriate remuneration, fair working conditions, diversity, and training and development opportunities; trade union rights and freedom of assembly; guarantee of adequate product safety, including health protection; application of the same requirements to entities in the supply chain; inclusive projects and consideration of the interests of communities and social minorities |
| Governance (G) | Tax honesty; anti-corruption measures; sustainability management by the board; board remuneration based on sustainability criteria; the facilitation of whistle blowing; employee rights guarantees; data protection guarantees; information disclosure |

These aspects were already partly addressed in the EU directive on non-financial disclosure in 2014, acknowledging the importance of publishing businesses information on sustainability such as social and environmental factors, with a view to identifying sustainability risks and increasing investor and consumer trust. Actually, disclosure of non-financial information is vital for managing change towards a sustainable global economy by combining long-term profitability with social justice and environmental protection. Thus, disclosure of non-financial information helps the measuring, monitoring and managing of undertakings' performance and their impact on society in order to take account of the multidimensional nature of corporate social responsibility (CSR) and the diversity of the CSR policies implemented by businesses matched by a sufficient level of comparability to meet the needs of investors and other stakeholders, as well as the need to provide consumers with easy access to information on the impact of businesses on society [35].

## 5. The role of small and medium sized insurance enterprises in Europe

Relatively little is, in general, known about CSR and ESG amongst small and medium sized enterprises (SMEs). Where SMEs are doing it, they may not use the language of CSR or ESG. Just as the best of SMEs are a source of innovation for business generally, so it can be assumed that the best of environmentally and socially responsible SMEs will offer CSR innovations. Efforts to engage more SMEs in CSR should be mindful of this fact. They should reflect the daily realities of SME life. They should work through channels as close to SMEs as possible, which SMEs already use and trust. This will involve a range of initiatives at local, regional, national, EU and sectoral levels. They will include initiatives from different stakeholder groups such as staff and consumers [36]. This is of special importance since the share of SMEs in total in the world is over 96%. This ratio is 99% in Germany, Japan, and France. Therefore, the place of SMEs in the economy is very important in general. It is known that SMEs play an important role in helping economic and social developments of countries as they rapidly adapt to the changing market conditions, have flexible production structures, achieve balanced growth among the regions, and help reduce unemployment [18]. In the insurance sector, typical SMEs are mutual insurance companies that reflect the original idea of insurance at best. Many of them were already founded in the late middle and early modern age as guilds or friendly societies in the Netherlands, Germany, France, England, and later, also in Northern America [37]. These institutions probably reflect the original insurance idea of a humanitarian protection against life or business existence threats at its best. In Germany, 241 out of a total of 535 insurance companies were organized in the form of a mutual company in 2019, which corresponds to 45% in number, although their share in the total premium income was only 14.7% [38].

With respect to ESG criteria, mutual SMEs can play an important role without much effort. We discuss several suggestions in more detail in the following section.

**Environmental:**

A careful asset management can concentrate on investments that are veritably Taxonomy-aligned ("green assets"). However, there is a delicate balancing act between risky assets which require additional solvency capital according to Solvency II (like investments in alternative energy supply) and less risky assets, which are probably not, or only in part, Taxonomy-aligned (like government and related bonds).

A new initiative that has come up recently is a direct sponsorship of local environmental projects (e.g. planting trees in the company's environment) or the establishment of a "green" non-profit foundation that provides financial resources for various local and cross-regional programmes that are not only restricted to environmental aspects but also to SG. Such a foundation has been incorporated recently by the GVO Mutual Insurance Company. The idea here is to donate regularly a certain monetary amount (say 1 Euro) per contract and year to the foundation. Similar activities can also apply to the company's head office building (green roofing, CO2 neutral energy supply, an improved heat insulation and other architectural aspects), the use of local supply chains for stationery and other office supplies, or an environmentally friendly car fleet.

But also, innovative insurance products can contribute to environmental protection and sustainability ("green insurance"). For instance, concerning the household contents insurance, the policy could guarantee a replacement of damaged technical devices with corresponding devices of the highest available energy efficiency class, e.g. refrigerators, washing machines, stoves and other appliances. In Germany, it has turned out that such a kind of insurance products is appealing more and more to the younger generation who has a higher mental affinity to environmental and climate protection. Seemingly, this clientele is also willing to pay a slightly higher premium in the awareness of a constructive contribution to environmental and climate protection.

Similar reflections also apply to agricultural insurance products, which are for a large part, handled by mutual insurance companies. The idea here is a kind of premium gratification system for countrymen who, e.g., avoid excessive chemical fertilizers or who care responsibly about their livestock breeding.

A completely new generation of insurance products concerns the private traffic sector. Due to new technologies like blockchain [39], it becomes possible to create car liability insurance products where the premium depends on the individual driving behaviour ("pay how you drive") [40]. This could give incentives to car owners to adapt their driving behaviour to the environment (less fuel consumption, less deterioration). Even if insurance SMEs are, in most cases, not able to handle the technical challenges connected with blockchain products themselves, they can easily serve as insurance brokers. A new line of insurance products emerges actually with bicycle insurance comprehensive coverage in Germany, in particular for pedelecs and e-bikes. This might give incentives to people to abstain from using cars with combustion engines in cities in favour of environment-friendly mobility.

**Social:**

SMEs are in general frequently characterized by flat management hierarchies. This facilitates to a great extent precaution measures for their employees, guarantees of equal opportunities and of workforce diversity, safety of workplaces, respecting a worker's council and the implementation of a staff unit for conflict management and whistle blowing.

On the personal side, SMEs can provide individual retirement arrangements, gratification programmes, support of continuing business education and honorary appointments, the establishment of an appropriate in-house television network for the employees, or in-house sportive activities within a general health programme. The GVO Mutual Insurance Company, for instance, sponsors general all sportive activities of their staff in-house and outdoor.

Social aspects are also immanent in many insurance products like life, accident, health, business interruption, and complementary insurance contracts.

**Governance:**

Although several legislative regulations prescribed by the European Taxonomy and Transparency Directive concern only large companies say with more than 500 employees, insurance SMEs can of course decide for a voluntary disclosure of their ESG activities, in particular when they pursue outstanding environmental sponsoring programmes, as e.g. the GVO Mutual Insurance Company does. Personally, a typically small and responsible Board of Directors of an SME can serve as business ideal for the employees, thus preventing stimulation of misconduct and deception. Further, sustainability aspects can and must become an indispensable part of the whole business culture, which, in particular, insurance SMEs can easily implement due to their flat business hierarchy. This concerns, besides business processes and service regulations supervised by the CEOs, the complete in-house staff as well as the employed field staff or sales department and the IT department of the company.

It is important to state that an implementation of all of the aforementioned examples will in general not lead to an increase in the company's risk profile which would be crucial in the light of the Solvency II directive.

## Conclusions

The world is rapidly changing due to an increase in climate variation and environment pollution, but also with respect to social problems like poverty, suppression and migration. In the awareness of these challenges, the UN has formulated several Sustainable Development Goals to be reached by 2030. As a reaction, the finance and insurance sector has initiated several activities to help overcoming these problems, accompanied by strong legal directives, particularly in Europe. A very important role here is played by insurance SMEs, which represent a significant number of companies in the insurance industry. Due to their flat management hierarchies and local business orientation, they are able to comply with almost all legal ESG demands without accumulating a higher risk profile, and can easily serve as forerunners in the propagation of ESG principles in the society. Thus, insurance SMEs can and will contribute to a promising way to reach the UN Sustainable Development Goals in the future.

## Acknowledgments

We would like to thank the Board of Directors of the GVO Mutual Insurance Company for permanent support and many fruitful discussions.

## Conflict of Interest

The authors declare no conflict of interest.